\documentclass{PoS}

\title{Prospect for observing a light charged Higgs through the decay $H^+ \to c \bar{b} + c.c.$ at the  LHeC }

\ShortTitle{Decay $H^+ \to c \bar{b} + c.c.$ at the LHeC }

\author{\speaker{J. Hern\'andez-S\'anchez}\thanks{Supported by SNI-CONACYT and PROMEP-SEP grants.}\\
        Facultad de Ciencias de la Electr\'onica, Benem\'erita Universidad Aunt\'onoma de Puebla, Apdo. Postal 542, C.P. 72570 Puebla, Puebla, M\'exico.\\
 Dual C-P Institute of High Energy Physics, Puebla, Pue., M\'exico.\\
        E-mail: \email{jaime.hernandez@correo.buap.mx}}

\author{O. Flores-S\'anchez\\
      Departamento de Sistemas y Computaci\'on
Instituto Tecnol\'ogico de Puebla,
Av. Tecnol\'ogico num. 420 Col. Maravillas,	Puebla, Puebla, C.P. 72220, M\'exico.\\
        E-mail: \email{omar.flores@itpuebla.edu.mx}}

\author{C. G. Honorato\\Facultad de Ciencias de la Electr\'onica, Benem\'erita Universidad Aunt\'onoma de Puebla, Apdo. Postal 542, C.P. 72570 Puebla, Puebla, M\'exico.\\
         E-mail: \email{carlosg.honorato@correo.buap.mx}}

\author{S. Moretti \\
School of Physics and Astronomy, University of Southampton, Highfield, Southampton SO17 1BJ, United Kingdom and Particle Physics Department, Rutherford Appleton Laboratory, Chilton, Didcot, Oxon OX11 0QX, United Kingdom \\
E-mail: \email{s.moretti@soton.ac.uk}} 
       
\author{S. Rosado\\
        Facultad de Ciencias F\'{\i}sico-Matem\'aticas,Benem\'erita Universidad Aut\'onoma de Puebla, Apdo. Postal 1364, C.P. 72570 Puebla, Puebla, M\'exico.\\
        E-mail: \email{sebastian.rosado@gmail.com}}

\abstract{We study the production and decay  of a light charged Higgs boson at the future Large Hadron electron Collider (LHeC) in the framework of the Two Higgs Doublet Type III, assuming a four-zero texture in the Yukawa matrices and a general Higgs potential.  We analyze the charge current production processes $e^- p \to \nu_e q H^+ $ considering the signature $H^+ \to c \bar{b} + c.c.$ of the final state.  We take this signature and we compare it  to the irreducible background from Standard Model (SM) interactions. We consider scenarios  of the model  which are consistent with current experimental data from Higgs and flavor physics. }

\FullConference{Prospects for Charged Higgs Discovery at Colliders\\
		3-6 October 2016\\
		Uppsala, Sweden}

\begin{document}
\section{Introduction}
Once discovered, a neutral Higgs boson at the Large Hadron Collider (LHC), by the ATLAS \cite{Atlas} and CMS \cite{cms} experiments, the SM is well established.  However, the SM-like limit exists for several extensions of the Higgs sector, e.g., the 
2-Higgs Doublet Model (2HDM) in its versions Type I, II, III (or Y)   and IV (or X), wherein Flavor Changing Neutral Currents (FCNCs) mediated by (pseudo)scalars can be eliminated under discrete symmetries \cite{Branco:2011iw}. Another, very interesting kind of  2HDM is  the one where FCFNs can be controlled by a particular texture in the Yukawa matrices  \cite{Branco:2011iw}. In particular, we have implemented a four zero-texture in a scenario which we have called 2HDM Type III (2HDM-III). This model has a phenomenology which is very rich, which we studied at colliders some years ago \cite{HernandezSanchez:2012eg,DiazCruz:2009ek},
 and some very interesting aspects, like flavor-violating quarks decays $\phi \to b \bar{s}$  ($\phi=h, \, H$), which can be enhanced for  neutral Higgs bosons with intermediate mass (i.e., below  the top quark mass) \cite{Das:2015kea}. Similarly, in this model, the parameter space can avoid the current experimental constraints of flavor  and Higgs physics and   a light charged Higgs  boson is allowed \cite{HernandezSanchez:2012eg}, so that  the  decay $H^+ \to c \bar{b}$ is enhanced and its Branching Ratio
 ($BR$) could be dominant. In fact, this channel  has been also studied in a variety of Multi-Higgs Doublet Models 
(MHDMs) \cite{Akeroyd:2016ymd,Akeroyd:2012yg}, wherein the $BR (H^+ \to c \bar{b}) \approx 0.7- 0.8$ and could afford one with a considerable gain in sensitivity to the presence of a 
$H^\pm$  by tagging the $b$ quark. In this work, we focus on the feasibility of finding at the future LHeC the charged Higgs boson of the 2HDM-III.

\section{The Higgs-Yukawa sector in the 2HDM-III}
 As we have shown previously, in the 2HDM-II,  when the four-zero-texture is implemented and FCNCs are controlled,   the discrete symmetry in the model is not necessary. Then the most general $SU(2)_L \times U(1)_Y$ invariant scalar potential for two scalar doublets, $\Phi_i^\dagger= (\phi_i^-, \phi_i^{0*})$ ($i=1, \, 2$),  must be considered, which is: 
\begin{eqnarray}
V(\Phi_1,\Phi_2)&=&\mu_1^2(\Phi_1^\dagger \Phi_1)+\mu_2^2(\Phi_2^\dagger \Phi_2)-\Big(\mu_{12}^2(\Phi_1^\dagger \Phi_2)+h.c.\Big)+\frac{1}{2}\lambda_1(\Phi_1^\dagger \Phi_1)^2+\frac{1}{2}\lambda_2(\Phi_2^\dagger \Phi_2)^2\nonumber\\
&&+\lambda_3(\Phi_1^\dagger \Phi_1)(\Phi_2^\dagger \Phi_2)+\lambda_4(\Phi_1^\dagger \Phi_2)(\Phi_2^\dagger \Phi_1)+\left[\frac{1}{2}\lambda_5(\Phi_1^\dagger \Phi_2)^2+\Big(\lambda_6(\Phi_1^\dagger \Phi_1)\right.\nonumber\\
&&\left.+\lambda_7(\Phi_2^\dagger \Phi_2)\Big)(\Phi_1^\dagger \Phi_2)+h.c.\right], \label{potencial}
\end{eqnarray}
where all parameters of the Higgs potential are assumed to be real, including the Vacuum Expectation Values (VEVs) of the scalar fields (i.e., the CP violation case is not considered). Hence, in the 2HDM-III, both doublets are coupled with the same fermions, so the Yukawa Lagrangian is given by:
\begin{equation}\label{yukawa-lagran}
{\cal L}_Y=-\Big(Y_1^u\bar Q_L \tilde \Phi_1 u_R+Y_2^u \bar Q_L\tilde \Phi_2 u_R+Y_1^d\bar Q_L \Phi_1 d_R+Y_2^d\bar Q_L \Phi_2 d_R+Y_1^l\bar L \Phi_1 l_R+Y_2^l \bar L_L \Phi_2 l_R\Big),
\end{equation}
where $\tilde \Phi_{1,2}= i \sigma_2 \Phi_{1,2}^*$. Since, in this equation,  the fermion mass matrices after  
Electro-Weak  Symmetry  Breaking (EWSB) are expressed by: 
$M_f = \frac{1}{2} (v_1 Y_1^f + v_2 Y_2^f )$, $f=u, \, d, \, l$, assuming that both Yukawa matrices  $Y_1$ and $Y_2$ have the four zero-texture form  and are Hermitian \cite{Fritzsch:2002ga}. Once we diagonalize the mass matrices, one can get the rotated matrix $Y_n^f $ with the following generic  expression:
\begin{equation}\label{Matrix-Y}
[\tilde Y_n^f]_{ij}=\frac{\sqrt{m_i^fm_j^f}}{v}[\tilde\chi_n^f]_{ij}=\frac{\sqrt{m_i^fm_j^f}} {v}[\chi_n^f]_{ij}e^{i\theta_{ij}^f},
\end{equation}
where the $\chi's$ are unknown dimensionless parameters of the model, with $v= \sqrt{v_1^2+v_2^2}$. Following the procedure of Ref. \cite{HernandezSanchez:2012eg}, we can obtain a generic expression for the couplings of the Higgs bosons with the fermions, by means of
\begin{eqnarray}
{\cal L}^{\bar f_i f_j \phi}&=&-\left\{\frac{\sqrt{2}}{v}\bar u_i(m_{d_j}X_{ij}P_R+m_{u_i}Y_{ij}P_l)d_jH^++\frac{\sqrt{2}m_{l_j}}{v}Z_{ij}\bar{v_L}l_RH^++h.c.\right\} \nonumber\\
&&-\frac{1}{v}\Big\{\bar f_im_{f_i}h_{ij}^ff_jh^0+\bar f_i m_{f_i}H_{ij}^ff_j H^0-i\bar f_im_{f_i}A_{ij}^f f_j \gamma_5 A^0 \Big\}, \label{Yukawa-Charged}
\end{eqnarray}
 where $\phi_{ij}^f$ ($\phi = h$, $H$, $A$), $X_{ij}$, $Y_{ij}$, $Z_{ij}$ are defined in \cite{HernandezSanchez:2012eg}, a structure that can be  related to  various incarnations of the 2HDM with  additional flavor physics in the Yukawa matrices, in such a way that the Higgs-fermion-antifermion coupling can be written as $g^{\phi ff}_{\rm 2HDM-III} = g^{\phi ff}_{\rm 2HDM-any} + \Delta g $, where $g^{\phi ff}_{\rm 2HDM-any}$ is the coupling $\phi f\bar f$  in  2HDMs with a $Z_2$ discrete symmetry and $\Delta g$ is the contribution of the four-zero texture. 

Now, we consider three different realisations  of the 2HDM-III, which give an   enhancement to the  decay
channel  $H^+ \to  c \bar{b} + h.c.$, in fact, this   $BR$ could be the leading one. 

\section{Benchmark scenarios of the 2HDM-III} 
We perform a parameter scan of the 2HDM-III similar to that of Ref. \cite{Das:2015kea}, where all recent experimental bounds from flavor  and Higgs physics (including EW Precision Observables (EWPOs))
were taken into account, alongside perturbativity, vacuum stability and unitarity constraints   \cite{HernandezSanchez:2012eg, Cordero}. Then, we choose the same benchmark scenarios given in \cite{Das:2015kea}, because in these cases the charged Higgs mass is light (110 GeV $\leq m_{H^\pm}\leq 200$ GeV). From the parameter space that survives current theoretical requirements and experimental data, we take the points that present a substantial enhancement of the decay $H^+ \to c \bar{b} + h.c.$, whose $BR(H^+ \to c \bar{b} + h.c.) \approx 1 $. Thus, the following benchkmark scenarios are considered:
\begin{itemize}
\item {\bf Scenario Ia}: $\cos(\beta-\alpha)=0.1$, $\chi^u_{kk}=1.5$ ($k=2$, 3),  $\chi_{22}^d =1.8$, $\chi_{33}^d =1.2$, $\chi_{23}^{u,d}=0.2$, $\chi_{22}^\ell=0.5$, $\chi_{33}^\ell=1.2$, $\chi_{23}^\ell=0.1$, $M_A=100 $ GeV, $m_{H^\pm}=$ 110 GeV, taking $Y = -X =-Z= \cot \beta$.
\item {\bf Scenario Ib}: the same of Scenario Ia, but with $\cos(\beta-\alpha)=0.5$.
\item {\bf Scenario IIa}: $\cos(\beta-\alpha)=0.1$, $\chi^u_{22}=0.5$, $\chi^u_{33}=1.4$,  $\chi_{22}^d =2$, $\chi_{33}^d =1.3$, $\chi_{23}^{u}=-0.53$, $\chi_{23}^{d}=0.2$, $\chi_{22}^\ell=0.4$, $\chi_{33}^\ell=1.2$, $\chi_{23}^\ell=0.1$, $M_A=100 $ GeV, $m_{H^\pm}=$ 110 GeV, taking $X =Z= \tan \beta=1/Y$.
\item {\bf Scenario Ya}: $\cos(\beta-\alpha)=0.1$, $\chi^u_{22}=0.5$, $\chi^u_{33}=1.4$,  $\chi_{22}^d =2$, $\chi_{33}^d =1.3$, $\chi_{23}^{u}=-0.53$, $\chi_{23}^{d}=0.2$, $\chi_{22}^\ell=0.4$, $\chi_{33}^\ell=1.1$, $\chi_{23}^\ell=0.1$, $M_A=100 $ GeV, $m_{H^\pm}=$ 110 GeV, taking $X =1/Y= \tan \beta=-1/Z$.
\end{itemize}

\section{Numerical analysis}
\begin{figure}[ht!]
\centering
 \includegraphics[width=2.9in]{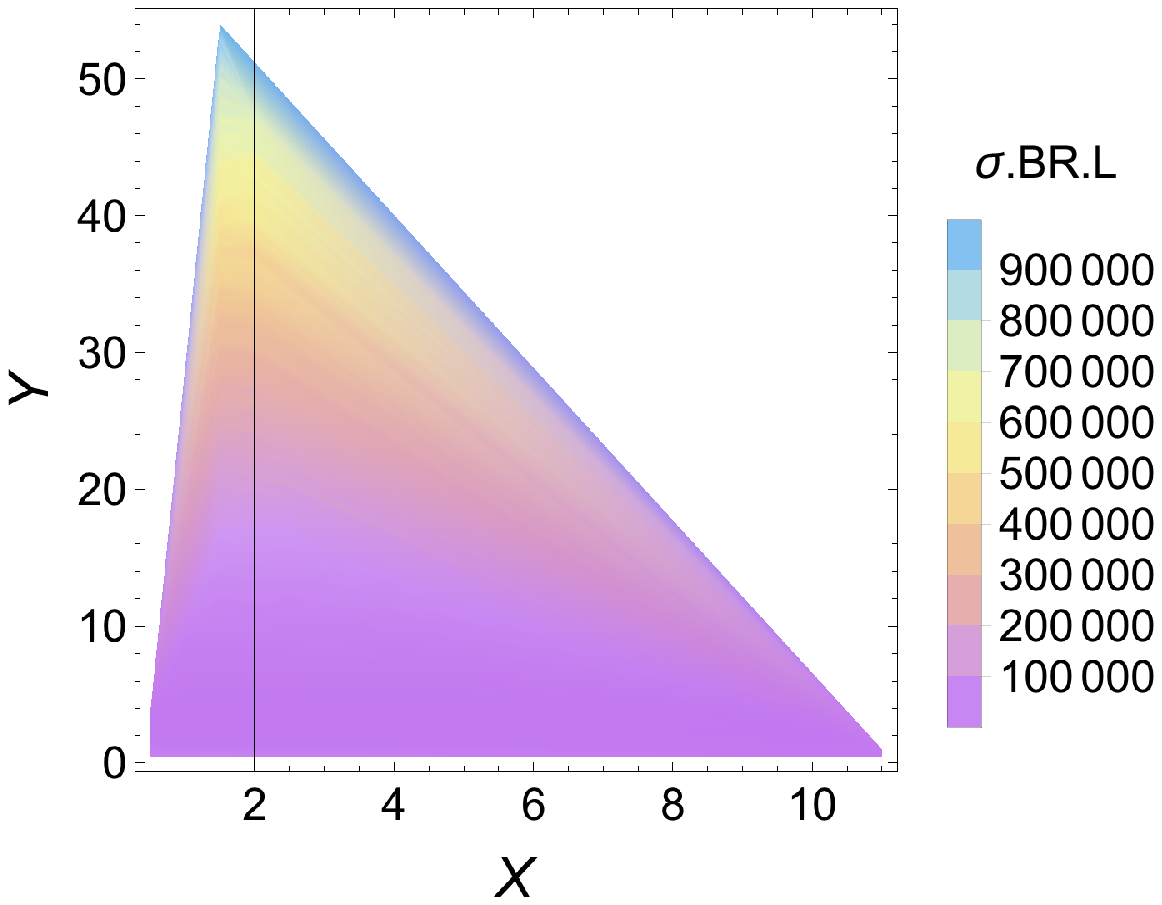}
 \includegraphics[width=2.9in]{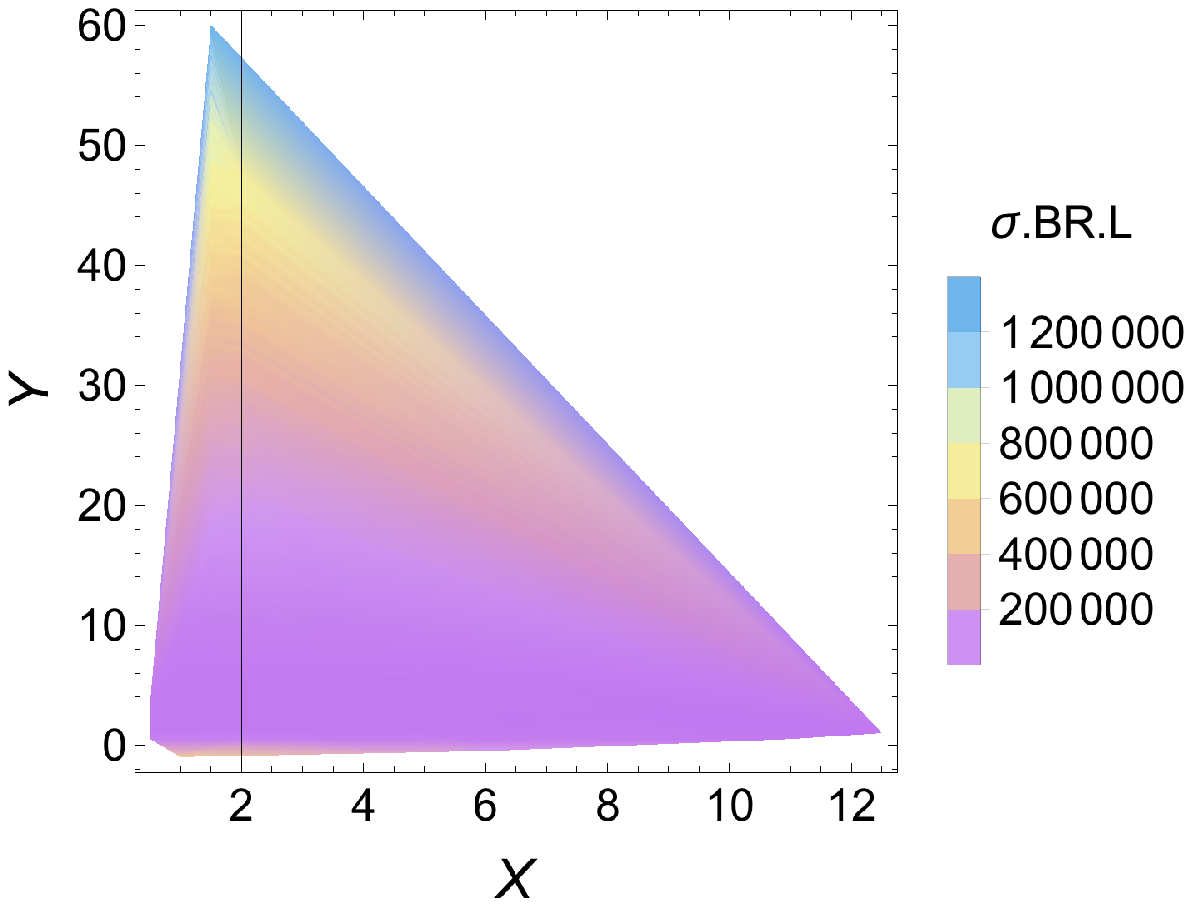}
\caption{The allowed region in the plane $X$ vs $Y$, using the constraint  from the radiative inclusive decay $B\to X_s \gamma$. We show the event rates  $(\sigma.BR. L)$ at parton level   for Scenario Ia (left panel) and  Scenario Ib (right panel), where $L$ is the integrated luminosity. We show both scenarios for $ L=100$ fb$^{-1}$ and taking $m_{H^\pm}=110$ GeV. }
\label{BPI}
\end{figure}

\begin{figure}[ht!]
\centering
 \includegraphics[width=2.9in]{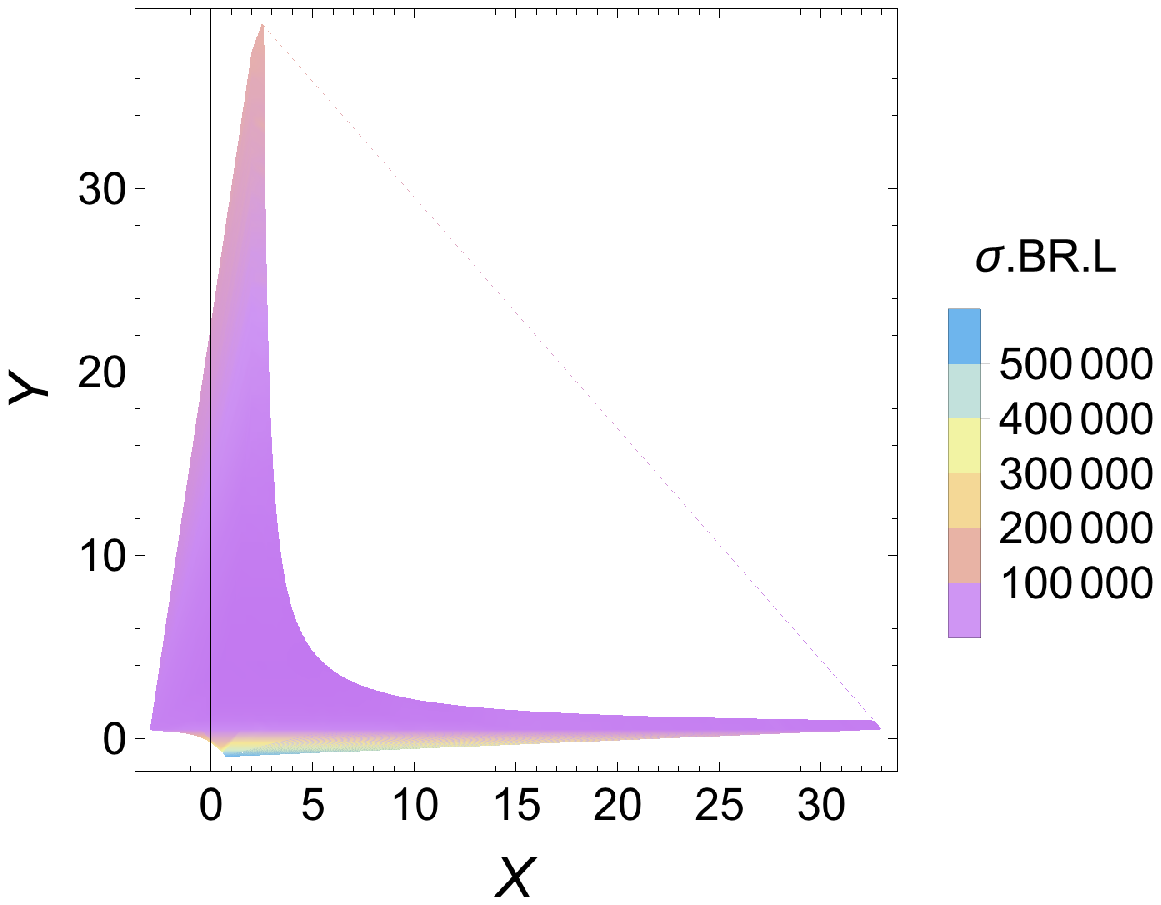}
 \includegraphics[width=2.9in]{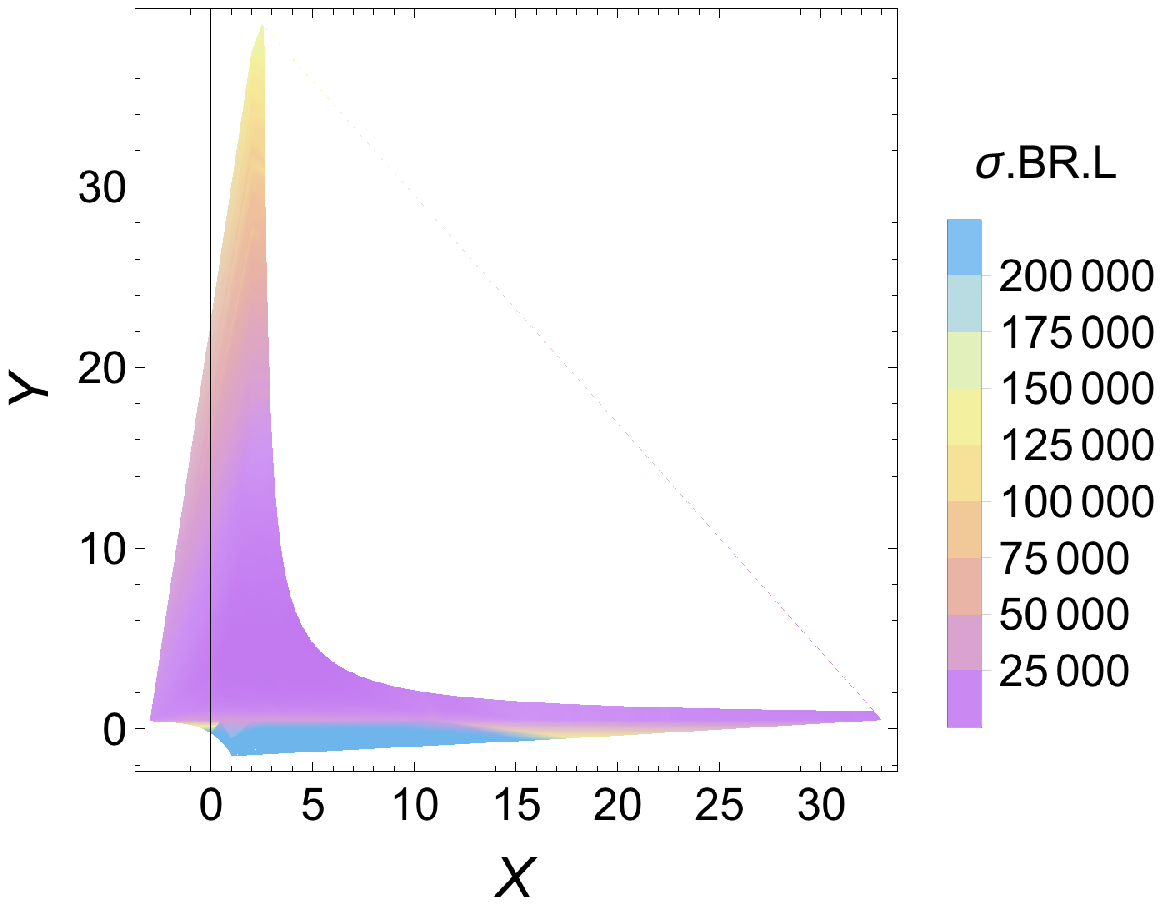}
\caption{The same as the previous plot    for  Scenario IIa (left panel) and Scenario Ya (right panel).}
\label{BPII}
\end{figure}

We study the process $e^- p \to \nu_e H^- b$ proceeding via two subprocesses: $e^- \bar{b} \to \nu_e \bar{t} \to \nu_e H^- \bar{b} $ (single top quark production and decay) and $e^- \bar{b} \to \nu_e W^{\pm *} \phi^* \to \nu_e H^- \bar{b} $ (vector-scalar fusion) \cite{Moretti:1997ip}. We assume that $H^- \to \bar{c} b$ is dominant. Our signal contains three jets (one is forward and two are central), missing transverse energy and no lepton. Out of the two central jets, one is $b$-tagged. We estimate the parton level signal cross section and $BR (H^+ \to c \bar{b} + h.c.) $ using CalcHEP \cite{Belyaev:2012qa}. For this study at the LHeC, we consider an electron beam with energy $E_{e^-}=60$ GeV while the energy of the proton beam is $E_p=7000$ GeV, which correspond to a center-of mass energy $\sqrt{s} =1.3$ TeV. The integrated luminosity is 100 fb$^{-1}$. In order to estimate the event rates at parton level we implement the following basic preselections:
\begin{eqnarray}
p_T^q> 15 ~{\rm GeV}, \,\,\,\,\,\,\,\,\,\,\,\, \Delta R(q,q)> 0.4,
\label{pres}
\end{eqnarray}
with $\Delta R(q,q) = \Delta \eta^2 +\delta \phi^2 $, where $\eta$ and $\phi$ are the rapidity and azimuthal angle, respectively. Under these considerations, we calculate in our benchmark scenarios the events rates ($\sigma. BR. L$). In  Figs. \ref{BPI} and \ref{BPII}, we show the event rates ($\sigma.BR.L$) at parton level for a charged Higgs $H^\pm$ in  Scenarios Ia-Ib and IIa-Ya. respectively. One can see that blue regions contain the most optimistic benchmark points for all scenarios and the events rates are very spectacular numbers, at times well above
the background.  In Tab. \ref{Tab1}, one can see the products of cross section times the relevant BRs ($\sigma.cb$).

\section{Backgrounds}
There are two groups of SM backgrounds for our signal: the charged-current backgrounds, $\nu t \bar{b}$, $\nu b  \bar{b} j$, and  photo-production, $e^- b  \bar{b} j$, 
$e^- t  \bar{t} j$. In order to estimate the cross sections of these SM backgrounds, we have used the same preselection as in eq. (\ref{pres}). The expected numbers of events for 100 fb$^{-1}$ of integrated luminosity are given in  Ref. \cite{Das:2015kea}. Then, we ought to consider, still at parton level (the hadron level analysis is in progress), that the $b$-jets in  both signal and background can only be tagged with  probability $\epsilon_b=0.5$. In the same way, we also adopted mistagging of non-$b$ jets, i.e., treated  gluon/light-flavor jets
 as  well as $c$-jets
 with a probability of $\epsilon_j=0.01$ (for $j=u, \, d, \, s, \, g $) and $\epsilon_c=0.1$, respectively. With this information, we can apply the tagging probability $\epsilon_b^2.\epsilon_c$ to the signal $S$ and $\epsilon_b^2.\epsilon_j$ to the background $B$. Taking in account these probabilities, we can get the significance at parton level for our benchmark points, which are shown in Tab. \ref{Tab2}.   
With these results one can obtain a significance of 3--4 $\sigma$, with 100 fb$^{-1}$ of integrated luminosity for a charged Higgs mass $m_{H^\pm}= 110$ GeV, $X=Y=5$ in Scenario Ia and Ib. In fact, the same happens for  Scenarios IIa and Ya when $X=32$ and $Y=0.5$. 
\begin{table}
\begin{center}
\begin{scriptsize}
\begin{tabular}{|c|c|c|c|c|c|c|c|c|c|c|c|}
\hline
\hline
2HDM & $X$ & $Y$ & $Z$ & \multicolumn{2}{c|}{$m_H^{\pm} = 110 \ {\rm GeV}$}   \\
 &  &  &  & $cb$ & $\sigma . cb$   \\ \hline
\hline
Ia & 5 &5  & 5 & 0.99& 97.36   \\
Ib & 5 & 5 & 5 & 0.99 &  99.80  \\
\hline
IIa &  32& 0.5 & 32 & 0.99 &  92.00  \\
\hline
Ya & 32 &0.5  &0.5  & 0.99 &  75.12  \\
\hline
\hline
\end{tabular}
\end{scriptsize}
\caption{Parameters for a few optimistic benchmark points in the 2HDM-III as a 2HDM-I, -II and -Y configuration.
Here {\it cb} stands for BR($H^+ \to c \bar b + h.c.$) while $\sigma$.$cb$ stands for the
cross section multiplied by the above $BR$ as obtained at the LHeC  in units
of fb. We have analyzed only the benchmarks where $\sigma.cb$ is greater
than 0.15 fb, so that at least 15 events are produced  for 100 fb$^{-1}$.
}\label{Tab1}
\end{center}
\end{table}
\begin{table}
\begin{center}
(Here, $\epsilon_b =0.50, \epsilon_c =0.1\  {\rm and} \ \epsilon_j =0.01$, where $j=u,d,s,g$)
  \begin{tabular}{| c | c | c | c |}
    \hline
    \hline
             & $S$ & $B$ & $\mathcal{S}= S/B^{1/2}$ \\ \hline \hline
     Ia ($X=5,Y=5$) & $243.4$ & $3835.1$ & $3.9$ \\ \hline \hline
     Ib ($X=5,Y=5$) & $249.5$ & $3835.1$ & $4.0$ \\ \hline \hline
     II ($X=32,Y=0.5$) & $230$ & $3835.1$ & $3.7$ \\ \hline \hline
     Y  ($X=32,Y=0.5$) & $187.8$ & $3835.1$ & $3.0$ \\
    \hline
  \end{tabular}
  \caption{Significances after 100 fb$^{-1}$ for a few optimistic benchmark points in the 2HDM-III as a 2HDM-I, -II and -Y configuration.
Here we have considered at parton level the signal reduced by the factor $\epsilon_b^2.\epsilon_c$  while the background from the SM is scaled by $\epsilon_b^2.\epsilon_j$.}\label{Tab2}
\end{center}
\end{table}

\section{Conclusions}
At the future LHeC, with a integrated luminosity of 100 fb$^{-1}$, we found at parton level that a charged Higgs boson of the 2HDM-III would be observed with approximately a  3--4 $\sigma$ significance. At the end of the LHeC
era,   with $1000$ fb$^{-1}$ of data,  the detection of such a charged Higgs boson would be  certain.


\begin{thebibliography}{99}
\bibitem{Atlas} G. Aad et. al. [ATLAS Collaboration], "Observation of a new particle in the search for the Standard Model Higgs boson with the ATLAS detector at the LHC", Phys. Lett. B {\bf 716}, 1 (2012). 

\bibitem{cms} S. Chatchyan et. al. [CMS Colllaboration], "Observation of a new boson at mass of 125 GeV with the CMS experiment at the LHC", Phys. Lett. B {\bf 716}, 30 (2012).



\bibitem{Branco:2011iw} 
  G.~C.~Branco, P.~M.~Ferreira, L.~Lavoura, M.~N.~Rebelo, M.~Sher and J.~P.~Silva,
  Phys.\ Rept.\  {\bf 516}, 1 (2012)
  [arXiv:1106.0034 [hep-ph]].
  
  
\bibitem{HernandezSanchez:2012eg} 
  J.~Hernandez-Sanchez, S.~Moretti, R.~Noriega-Papaqui and A.~Rosado,
  JHEP {\bf 1307}, 044 (2013)










\bibitem{DiazCruz:2009ek} 
  J.~L.~Diaz-Cruz, J.~Hernandez--Sanchez, S.~Moretti, R.~Noriega-Papaqui and A.~Rosado,
  Phys.\ Rev.\ D {\bf 79}, 095025 (2009)
  [arXiv:0902.4490 [hep-ph]].

\bibitem{Das:2015kea} 
  S.~P.~Das, J.~Hernandez-Sanchez, S.~Moretti, A.~Rosado and R.~Xoxocotzi,
  Phys.\ Rev.\ D {\bf 94}, no. 5, 055003 (2016)
  [arXiv:1503.01464 [hep-ph]].


\bibitem{Akeroyd:2016ymd} 
  A.~G.~Akeroyd {\it et al.},
  arXiv:1607.01320 [hep-ph].



\bibitem{Akeroyd:2012yg} 
  A.~G.~Akeroyd, S.~Moretti and J.~Hernandez-Sanchez,
  Phys.\ Rev.\ D {\bf 85}, 115002 (2012)
  [arXiv:1203.5769 [hep-ph]].
  
\bibitem{Fritzsch:2002ga} 
  H.~Fritzsch and Z.~z.~Xing,
  Phys.\ Lett.\ B {\bf 555}, 63 (2003)
  [hep-ph/0212195].

\bibitem{Cordero}  A. Cordero-Cid et al. "Impact of a four zero Yukawa texture on $h \rightarrow \gamma\gamma$ and $\gamma Z$ in the framework of the Two Higgs Doublet Model Type III", JHEP {\bf 1407}, 057 (2014). 

\bibitem{Moretti:1997ip} 
  S.~Moretti and K.~Odagiri,
  Phys.\ Rev.\ D {\bf 57}, 5773 (1998)
  [hep-ph/9709458].

\bibitem{Belyaev:2012qa} 
  A.~Belyaev, N.~D.~Christensen and A.~Pukhov,
  ``CalcHEP 3.4 for collider physics within and beyond the Standard Model,''
  Comput.\ Phys.\ Commun.\  {\bf 184}, 1729 (2013).






\end{thebibliography}
\end{document}